\documentstyle[aps,12pt,amssymb,epsfig]{revtex}

\begin{document}

\baselineskip=0.60cm
\newcommand{\ini}{\begin{equation}}
\newcommand{\fin}{\end{equation}}
\newcommand{\inir}{\begin{eqnarray}}
\newcommand{\finr}{\end{eqnarray}}
\newcommand{\inif}{\begin{figure}}
\newcommand{\finf}{\end{figure}}
\newcommand{\bc}{\begin{center}}
\newcommand{\ec}{\end{center}}

\def\ol{\overline}
\def\pa{\partial}
\def\ra{\rightarrow}
\def\ts{\times}
\def\df{\dotfill}
\def\bs{\backslash}
\def\dg{\dagger}

$~$

\hfill DSF-15/2002

\vspace{1 cm}

\centerline{{\bf LEPTON NUMBER AND LEPTON FLAVOR VIOLATIONS}}
\centerline{{\bf IN SEESAW MODELS}}

\vspace{1 cm}

\centerline{\large{D. Falcone}}

\vspace{1 cm}

\centerline{Dipartimento di Scienze Fisiche, Universit\`a di Napoli,}
\centerline{Complesso di Monte S. Angelo, Via Cintia, Napoli, Italy}

\vspace{1 cm}

\begin{abstract}

\noindent
We discuss the impact of fermion mass matrices on some lepton number violating
processes, namely baryogenesis via leptogenesis and neutrinoless double beta
decay, and on some lepton flavor violating processes, namely radiative lepton
decays in supersymmetric seesaw models.

\end{abstract}

\newpage

\section{Introduction}

A breakthrough in particle physics happened in 1998, when the
SuperKamiokande Collaboration announced evidence for oscillation of
atmospheric neutrinos \cite{sk}. Recently, at the Sudbury Neutrino
Observatory (SNO), evidence for flavor conversion of solar neutrinos has been
found \cite{sno}, pointing towards oscillation of solar neutrinos too.
These two important results come after a long series of experiments.

The most natural explanation of neutrino oscillations is that neutrinos
have masses, and leptons mix just like quarks do. In this case, neutrino
mass eigenstates $\nu_i$ are related to neutrino flavor eigenstates
$\nu_{\alpha}$ by the unitary transformation 
$\nu_{\alpha}=U_{\alpha i} \nu_i$, where $U$ is the lepton mixing matrix
\cite{mns}.

It turns out that neutrino masses are very small with respect to charged
lepton and quark masses. This fact can be accounted for in a simple and
elegant way by means of the seesaw mechanism \cite{ss}, which requires only
a modest extension of the minimal standard model, namely the addition
of the right-handed neutrinos.

As a consequence of this inclusion, a Yukawa term generating a Dirac mass term
for the neutrino is allowed. Moreover, a Majorana mass term for the
right-handed neutrino is also allowed. While the Dirac mass, $m_{\nu}$,
is expected to be of the same order of magnitude as the quark or charged
lepton mass, the Majorana mass of the right-handed neutrino, $m_R$, is
not constrained and thus may be very large. If this case occurs, a small
effective Majorana mass for the left-handed neutrino,
$m_L \simeq (m_{\nu}/m_R) m_{\nu}$, is generated.

In this framework, lepton flavors and lepton number are not conserved.
The amount of lepton flavor violations may be very different, according
to supersymmetric (SUSY) or nonsupersymmetric (nonSUSY) models.
In fact, in nonSUSY models, due to the smallness of neutrino masses,
lepton flavor violations are so small to be unobservable \cite{petcov}.
Instead, in
SUSY models with universal soft breaking terms, lepton flavor violations
can get much enhanced with respect to nonSUSY models \cite{bm}.

On the other hand, the seesaw mechanism allows for lepton number violations,
such as the neutrinoless double beta decay and the right-handed neutrino decay.
The latter may be involved in the generation of the baryon asymmetry in
the universe, through the baryogenesis via leptogenesis mechanism \cite{fy}.
In fact, the lepton asymmetry produced by the decay of heavy right-handed
neutrinos is partially converted into a baryon asymmetry by electroweak sphaleron
processes \cite{krs}.

In the present paper, we discuss both lepton number and lepton flavor
violations in nonSUSY and SUSY seesaw models, especially in connection
with fermion mass matrices. We consider some explicit models for mass
matrices and determine the implications for the baryogenesis via leptogenesis,
the neutrinoless double beta decay and the radiative lepton decays in
SUSY theories.
In section II we summarize the experimental informations on neutrino masses
and lepton mixings. In section III the seesaw mechanism is briefly discussed.
In section IV and V we outline the link between mass matrices, leptogenesis
and radiative lepton decays in SUSY models. Finally, in section VI, we comment
on the results.

\section{Neutrino masses and mixings}

From the combined study of atmospheric, solar and also reactor neutrinos
we get a nearly bimaximal form for the mixing matrix,
\ini
U \simeq \left( \begin{array}{ccc}
\frac{1}{\sqrt2} & \frac{1}{\sqrt2} & \epsilon \\
-\frac{1}{2} & \frac{1}{2} & \frac{1}{\sqrt2} \\
\frac{1}{2} & -\frac{1}{2} & \frac{1}{\sqrt2}
\end{array} \right),
\fin
where $\epsilon \lesssim 0.1$. The name bimaximal comes from the fact that
both $U_{\mu 3}$ and $U_{e 2}$ have the value $1/\sqrt2$, while $U_{e 3}$
is very small. However, the best fit is closer to
\ini
U \simeq \left( \begin{array}{ccc}
\frac{2}{\sqrt6} & \frac{1}{\sqrt3} & \epsilon \\
-\frac{1}{\sqrt6} & \frac{1}{\sqrt3} & \frac{1}{\sqrt2} \\
\frac{1}{\sqrt6} & -\frac{1}{\sqrt3} & \frac{1}{\sqrt2}
\end{array} \right).
\fin
The element $U_{\mu 3}$ is determined by atmospheric oscillations, with
a mixing angle nearly maximal. The element $U_{e 2}$ by solar oscillations,
with a mixing angle large but not maximal. The smallness of element $U_{e 3}$
is obtained from reactor neutrinos \cite{chooz}.

These studies also provide values for square mass differences of left-handed
neutrinos. From atmospheric neutrinos we get
\ini
|m_3^2-m_2^2| \sim 10^{-3} \text{eV},
\fin
and from solar neutrinos (LMA solution)
\ini
|m_2^2-m_1^2| \sim 10^{-5} \text{eV},
\fin
or less favoured (LOW solution)
\ini
|m_2^2-m_1^2| \sim 10^{-7} \text{eV}.
\fin
Other two experimental informations come from single beta decay \cite{beta1},
\ini
m_{\nu_e}=(|U_{ei}|^2 m_i^2)^{1/2} < 2.5~ \text{eV},
\fin
and neutrinoless double beta decay \cite{beta2},
\ini
M_{ee}=U_{ei}^2 m_i \lesssim 0.38~ \text{eV}.
\fin
The process of neutrinoless double beta decay is allowed only if neutrinos
are Majorana particles, thus evidence for it would support the seesaw
mechanism.

Due to the property $|m_2^2-m_1^2|  \ll |m_3^2-m_2^2|$, three types of
hierarchies for left-handed neutrinos are possible. In the normal hierarchy
$m_3 \gg m_2,m_1$, with a complete hierarchy for $m_2 \gg m_1$. In this
case $m_3^2 \sim 10^{-3}$ eV$^2$ and $m_2^2 \sim 10^{-5}$ eV$^2$
(or $m_2^2 \sim 10^{-7}$ eV$^2$). The partially degenerate spectrum is obtained
for $m_2 \simeq m_1$. In the inverse hierarchy,
$m_1 \simeq m_2 \gg m_3$, we get $m_{1,2}^2 \sim 10^{-3}$ eV$^2$.
Finally, for the nearly degenerate spectrum we have 
$m_1 \simeq m_2 \simeq m_3 \sim 0.1-1$ eV, because of relations (6) and (7).

In general, the lepton mixing matrix can be parametrized as the standard form
of the quark mixing matrix (including a phase $\delta$),
times a diagonal phase matrix, like
$P=\text{diag}(\text{e}^{\text{i}\varphi_1/2},
\text{e}^{\text{i}\varphi_2/2},1)$.
Sometimes a simplified approach is useful, namely to consider $m_1,m_2$
to be positive and negative, neglecting phases $\varphi_1,\varphi_2$.
In a similar way one can take $\epsilon$ positive or negative, and neglect
the phase $\delta$. Negative masses $m_{1,2}$ correspond to phases
$\varphi_{1,2}=\pi$.

\section{Seesaw mechanism}

For three generations of fermions the seesaw formula is given by
\ini
M_L \simeq M_{\nu} M_R^{-1} M_{\nu}^T,
\fin
where $M_{\nu}$ is the Dirac neutrino mass matrix, $M_R$ the right-handed
neutrino mass matrix and $M_L$ the left-handed (effective) neutrino mass matrix.
Some problems with naturalness may happen. In fact, if $M_{\nu}$ is highly
hierarchical, as quark or charged lepton mass matrices are, then it is
unnatural to obtain nearly degenerate neutrinos.

From the experimental informations on neutrino masses and mixings we can
infer the possible forms of the effective neutrino mass matrix through
the formula
\ini
M_L = U D_L U^T,
\fin
where $D_L$ is the diagonal of effective neutrino masses. This relation is
valid in the basis with the charged lepton mass matrix diagonal, $M_e=D_e$.
However, for $M_e$ nearly diagonal, the approximation (9) can be adopted
because of the bilarge lepton mixing. In fact, the difference between matrices
(1) and (2) could be due to the contribution of the charged lepton mass matrix
\cite{gt} or to a running effect from a high scale \cite{aklr}.
We are interested in determining the form
of the heavy neutrino mass matrix through the inverse seesaw formula
\ini
M_R \simeq M_{\nu}^T M_L^{-1} M_{\nu}.
\fin
Note that the matrix $M_L^{-1}$ can be obtained from $M_L$ by changing $m_i$
with $1/m_i$, because $M_L^{-1}=U D_L^{-1} U^T$.
As a first step we may assume symmetric matrices and quark-lepton symmetry,
\ini
M_e \sim M_d \sim \text{diag}(m_d,m_s,m_b),
\fin
\ini
M_{\nu} \sim M_u \sim \text{diag}(m_u,m_c,m_t).
\fin
The first quark-lepton relation is indeed a good approximation,
while the second one is only an assumption.

\section{Lepton number violation}

If we consider the seesaw mechanism, we have light (left-handed, effective)
and heavy (right-handed) Majorana neutrinos and hence the violation of
total lepton number. The neutrinoless double beta decay is allowed,
with $M_{ee}=M_{L11}$. Moreover, as a consequence of electroweak sphaleron
processes, the lepton number violation can be converted into a baryon number
violation. Then, the baryogenesis via leptogenesis mechanism was proposed
\cite{fy,luty}
where the out-of-equilibrium decays of heavy neutrinos produce a lepton
asymmetry which is tranformed into a baryon asymmetry by sphaleron
processes.

In the baryogenesis via leptogenesis mechanism, the baryon asymmetry is given by
\ini
Y_B \simeq \frac{1}{2}~\frac{1}{g^*}~d~\epsilon_1,
\fin
with the CP violating asymmetry in the decay of the lightest heavy neutrino
with mass $M_1 \ll M_2 < M_3$ given by
\ini
\epsilon_1 \simeq \frac{3}{16 \pi v^2}
\left[\frac{[(M_D^{\dg} M_D)_{12}]^2}{(M_D^{\dg} M_D)_{11}}
\frac{M_1}{M_2}+\frac{[(M_D^{\dg} M_D)_{13}]^2}{(M_D^{\dg}
M_D)_{11}}
\frac{M_1}{M_3} \right],
\fin
where $M_D=U_e^{\dg} M_{\nu} U_R$, with $U_R$ diagonalizing $M_R$ and $U_e$
diagonalizing $M_e$. The parameter $v \simeq m_t$ is the VEV of the Higgs doublet.
The factor $d$ in (13) is a dilution factor which depends on $M_1$ and
especially on
\ini
\tilde{m}_1=\frac{(M_D^{\dg} M_D)_{11}}{M_1}.
\fin
Minor dilution, $d \sim 10^{-1}$ is achieved for 
$\tilde{m}_1=10^{-5}-10^{-2}$ eV, while outside this range the dilution factor
drops \cite{bp}. Primordial nucleosynthesis requires $Y_B$ to lie between
$10^{-11}$ and $10^{-10}$ (see for example Ref.\cite{olive}).  

We consider realistic mass matrices,
expressed in terms of the Cabibbo parameter $\lambda=0.22$
and the overall mass scale,
\ini
M_{e} \sim \left( \begin{array}{ccc}
\lambda^{6} & \lambda^3 & \lambda^{5} \\
\lambda^3 & \lambda^2 & \lambda^2 \\
\lambda^{5} & \lambda^2 & 1
\end{array} \right)m_{b},
\fin
\ini
M_{\nu} \sim \left( \begin{array}{ccc}
\lambda^{12} & \lambda^6 & \lambda^{10} \\
\lambda^6 & \lambda^4 & \lambda^4 \\
\lambda^{10} & \lambda^4 & 1
\end{array} \right)m_t,
\fin
based on both $U(2)$ horizontal symmetry and quark-lepton symmetry
(see Ref.\cite{fal} and references therein).

For the complete normal hierarchy (and also
the inverse hierarchy and the partially degenerate spectrum with $m_2>0$)
we get
\ini
M_{R} \sim \left( \begin{array}{ccc}
\lambda^{12} & \lambda^{10} & \lambda^{6} \\
\lambda^{10} & \lambda^8 & \lambda^4 \\
\lambda^{6} & \lambda^4 & 1
\end{array} \right) \frac{m_t^2}{m_1},
\fin
consistent with the $U(2)$ symmetry \cite{fal}, and
\ini
\epsilon_1 \sim \frac{3}{16 \pi}
\left( \frac{\lambda^{20}}{\lambda^{12}}~\lambda^4+
\frac{\lambda^{12}}{\lambda^{12}}~\lambda^{12} \right) \sim 10^{-10},
\fin
with $\tilde{m}_1 \sim m_1$, providing $Y_B \sim 10^{-14}$. The overall mass
scale of $M_R$ is larger than $10^{15}$ GeV, which is close to the unification
scale.

For the partially degenerate spectrum with
$m_2<0$ and $m_1 \simeq \epsilon m_3$, so that $(M_L^{-1})_{33} \sim 0$,
we have
\ini
M_{R} \sim \left( \begin{array}{ccc}
\lambda^{10} & \lambda^{6} & \lambda^{4} \\
\lambda^{6} & \lambda^4 & 1 \\
\lambda^{4} & 1 & \lambda^2
\end{array} \right) \lambda^6 ~\frac{m_t^2}{m_1},
\fin
\ini
\epsilon_1 \sim \frac{3}{16 \pi}
\left( \frac{\lambda^{16}}{\lambda^{12}}~\lambda^4+
\frac{\lambda^{12}}{\lambda^{12}}~\lambda^{4} \right) \sim 10^{-4},
\fin
with $\tilde{m}_1 \simeq \lambda^2 m_1$, so that high baryon asymmetry
is achieved. Here the overall mass scale of $M_R$ is larger than $10^{11}$ GeV,
close to an intermediate scale.

If a moderate hierarchy in $M_{\nu}$ is adopted, for example
\ini
M_{\nu} \sim \left( \begin{array}{ccc}
\lambda^{6} & \lambda^3 & \lambda^{5} \\
\lambda^3 & \lambda^2 & \lambda^2 \\
\lambda^{5} & \lambda^2 & 1
\end{array} \right)m_t,
\fin
we obtain
\ini
M_{R} \sim \left( \begin{array}{ccc}
\lambda^{6} & \lambda^{5} & \lambda^{3} \\
\lambda^{5} & \lambda^4 & \lambda^2 \\
\lambda^{3} & \lambda^2 & 1
\end{array} \right) \frac{m_t^2}{m_1},
\fin
\ini
\epsilon_1 \sim \frac{3}{16 \pi}
\left( \frac{\lambda^{10}}{\lambda^{6}}~\lambda^2+
\frac{\lambda^{6}}{\lambda^{6}}~\lambda^{6} \right) \sim 10^{-6},
\fin
with $\tilde{m}_1 \sim m_1$, providing $Y_B \sim 10^{-10}$. The overall mass
scale of $M_R$ is again larger than $10^{15}$ GeV,
close to the unification scale.

Thus we have considered three distinct models for lepton mass matrices \cite{falc}.
Model I is based on matrices (16), (17), (18), model II on matrices
(16), (17), (20), and model III on matrices (16), (22), (23).
Models I and III have $M_R$ nearly diagonal at the high scale,
while the model II has a roughly offdiagonal $M_R$ at the intermediate
scale. Models II and III are reliable for leptogenesis, while model I gives
a too small asymmetry. 
Moreover, due to the value $m_2 < 0$,
model II leads to a suppression of the rate for neutrinoless
double beta decay, $M_{ee} \sim 10^{-4}-10^{-3}$ eV, while models I and III
yield $M_{ee} \sim 10^{-3}-10^{-2}$ eV for the normal hierarchy and
$M_{ee} \sim 10^{-2}-10^{-1}$ eV for the inverse hierarchy.
The link between leptogenesis and lepton mass matrices is discussed in many
papers, see for example Ref.\cite{lep}.

\section{Lepton flavor violation}

In SUSY seesaw models with universality above the heavy neutrino
mass scale, lepton flavor violations are induced, which depend on the parameters
\ini
C_{ij} = \frac{1}{v^2}~(M_D^{\dg})_{ik}~ \text{ln} \frac{M_U}{M_k}~(M_D)_{kj},
\fin
where $M_U$ is the universality
scale. In fact, in SUSY models with soft breaking terms, there are lepton flavor
violating terms in the offdiagonal elements of slepton mass matrices and
trilinear couplings. If such violations occur at the tree level, the branching
ratios exceed the experimental bounds. Therefore, it is usually assumed
that lepton flavor violations do not occur at the tree level, and this is
realized by assuming universality, that is slepton mass matrices and
trilinear couplings proportional to the unit matrix, as happens
in minimal supergravity. However, lepton flavor violations are generated by the
effect of renormalization of Dirac neutrino Yukawa couplings from the universal
scale to the right-handed neutrino scale \cite{bm}. The offdiagonal elements
of the Dirac neutrino mass matrix induce offdiagonal elements in slepton
mass matrices and trilinear couplings. In particular, the branching ratios
for lepton flavor violating radiative processes $l_i \ra l_j + \gamma$, where $l$ stands
for a charged lepton and $\gamma$ for a photon, 
depend on the offdiagonal elements of slepton mass matrices, which
in turn depend on the quantities $C_{ij}$. The subject has been studied in
several papers, see for example Ref.\cite{papers}. Here we discuss the impact
of the mass matrices of the previous section on radiative lepton decays.
Note that the baryogenesis via leptogenesis and the radiative lepton decays
have a different dependence on the neutrino mass matrix $M_D$.

For model I we get
\ini
U_{e} \sim \left( \begin{array}{ccc}
1 & \lambda & \lambda^{5} \\
-\lambda & 1 & \lambda^2 \\
\lambda^{5} & -\lambda^2 & 1
\end{array} \right),
~~
U_{R} \sim \left( \begin{array}{ccc}
1 & \lambda^2 & \lambda^{6} \\
-\lambda^2 & 1 & \lambda^4 \\
\lambda^{6} & -\lambda^4 & 1
\end{array} \right),
\fin
so that
\ini
M_D=U_e^{\dg} M_{\nu} U_R \sim
\left( \begin{array}{ccc}
\lambda^{7} & \lambda^{5} & \lambda^{5} \\
\lambda^{6} & \lambda^4 & \lambda^2 \\
\lambda^{6} & \lambda^4 & 1
\end{array} \right) m_t.
\fin
We assume that cancellations between terms of the same order in $\lambda$
do not occur. The calculation of $C_{ij}$ gives
\ini
C_{12} \sim \lambda^{12} \ln \lambda^{12}  +
\lambda^{10} \ln \lambda^{8}+ \lambda^{10} \sim 10^{-7},
\fin
\ini
C_{23} \sim  \lambda^{10} \ln \lambda^{12}+
\lambda^{6} \ln \lambda^{8}+ \lambda^{4} \sim 10^{-3},
\fin
\ini
C_{13} \sim \lambda^{12} \ln \lambda^{12}+
\lambda^{8} \ln \lambda^{8}+ \lambda^{6} \sim 10^{-4}.
\fin
For model II we have
\ini
U_{R} \sim \left( \begin{array}{ccc}
1 & \lambda^4 & \lambda^{6} \\
-\lambda^4 & 1 & \lambda^2 \\
\lambda^{6} & -\lambda^2 & 1
\end{array} \right),
\fin
and
\ini
M_D \sim
\left( \begin{array}{ccc}
\lambda^{7} & \lambda^{5} & \lambda^{5} \\
\lambda^{6} & \lambda^4 & \lambda^2 \\
\lambda^{6} & \lambda^2 & 1
\end{array} \right) m_t,
\fin
which differs from matrix (27) only for the element 3-2, so that
\ini
C_{12} \sim  \lambda^{12} \ln \lambda^{4}+
\lambda^{10} \ln \lambda^{4}+ \lambda^{8} \sim 10^{-6},
\fin
\ini
C_{23} \sim  \lambda^{10} \ln \lambda^{4}+
\lambda^{6} \ln \lambda^{4}+ \lambda^{2} \sim 10^{-2},
\fin
\ini
C_{13} \sim  \lambda^{12} \ln \lambda^{4}+
\lambda^{8} \ln \lambda^{4}+ \lambda^{6} \sim 10^{-4}.
\fin
For model III we obtain
\ini
U_{R} \sim \left( \begin{array}{ccc}
1 & \lambda & \lambda^{3} \\
-\lambda & 1 & \lambda^2 \\
\lambda^{3} & -\lambda^2 & 1
\end{array} \right),
\fin
\ini
M_D \sim
\left( \begin{array}{ccc}
\lambda^{4} & \lambda^{3} & \lambda^{3} \\
\lambda^{3} & \lambda^2 & \lambda^2 \\
\lambda^{3} & \lambda^2 & 1
\end{array} \right) m_t,
\fin
and then
\ini
C_{12} \sim  \lambda^{7} \ln \lambda^{6}+
\lambda^{5} \ln \lambda^{4}+ \lambda^{5} \sim 10^{-3},
\fin
\ini
C_{23} \sim  \lambda^{6} \ln \lambda^{6}+
\lambda^{4} \ln \lambda^{4}+ \lambda^{2} \sim 10^{-2},
\fin
\ini
C_{13} \sim  \lambda^{7} \ln \lambda^{6}+
\lambda^{5} \ln \lambda^{4}+ \lambda^{3} \sim 10^{-2}.
\fin
Generally, the dominant term is the third. However, sometimes the dominant term
can be the second. Of course, each term has a coefficient of order 1 not
indicated.

We have assumed that the universality scale is larger but of the same order
of the heaviest right-handed neutrino mass. If $M_U \sim M_P$, the Planck
mass, then $C_{ij}$ is enhanced by about one order of magnitude.
The experimental bounds on $C_{ij}$, inferred from Ref.\cite{lms}, are given by
$C_{12} \lesssim 10^{-3}-10^{-1}$,
$C_{23} \lesssim 10^{-1}-10^{2}$,
$C_{13} \lesssim 10^{-3}-10^{-1}$, for large $\tan \beta$.
Because of uncertainties in SUSY parameters, only wide ranges are available.
Future sensitivities for $C_{12}$ and $C_{23}$ are expected to be lowered
by one or two orders in next years. Due to theoretical and experimental
uncertainties, we cannot make definite predictions. However, it is worth
stressing that generally models favoured for leptogenesis predict higher
values for $C_{ij}$, so that a positive signal could be found.

\section{Conclusion}

We have performed on order-of-magnitude analysis of some lepton number
violating processes, namely baryogenesis via leptogenesis and neutrinoless
double beta decay, and some lepton flavor violating processes, namely
radiative lepton decays in SUSY models. Three distinct kinds of model for
mass matrices have been used. Generally,
when leptogenesis is enhanced, also the 
rate of lepton decays is higher. Then, if lepton decays are not found,
this would possibly
imply another mechanism for baryogenesis or another mechanism for SUSY
breaking, for example, instead of the gravity mediated SUSY breaking,
the gauge mediated SUSY breaking \cite{martin}.

$~$

We thank F. Buccella, F. Tramontano, G. Ricciardi, A. Della Selva
for discussions.

\end{document}